\documentclass[12pt]{article}

\usepackage{graphicx}
\usepackage{cite}

\begin{document}

\hoffset = -1truecm \voffset = -2truecm \baselineskip = 6 mm

\title{\bf Applications of a nonlinear evolution equation I: the parton distributions in the proton}

\author{
{\bf Xurong Chen}$^1$, {\bf Jianhong Ruan}$^2$, {\bf  Rong Wang}$^{1,3}$\\
{\bf Pengming Zhang}$^1$ and {\bf Wei Zhu}$^2$\footnote{Corresponding author, E-mail: weizhu@mail.ecnu.edu.cn}\\
\\
\normalsize $^1$Institute of Modern Physics, Chinese Academy of Sciences, Lanzhou 730000,
 P.R. China\\
\normalsize $^2$Department of Physics, East China Normal University,
Shanghai 200062, P.R. China \\
\normalsize $^3$ University of Chinese Academy of Sciences, Beijing, 100049, P.R. China
}

\date{}

\newpage

\maketitle

\vskip 3truecm

\begin{abstract}
\noindent The nonlinear DGLAP evolution equations with parton
recombination corrections are used to dynamically evaluate the
proton's parton distribution functions starting from a low scale
$\mu^2$, where the nucleon consists of valence quarks. We find that
the resulting negative nonlinear corrections can improve the
perturbative stability of the QCD evolution equation at low $Q^2$.
Our resulting parton distributions, with four free parameters, are
compatible with the existing databases. This approach provides a
powerful tool to connect the quark models of the hadron and various
non-perturbative effects at the scale $\mu^2$ with the measured
structure functions at the high scale $Q^2>> \mu^2$.
\end{abstract}

PACS number(s):13.60.Hb; 12.38.Bx

$keywords$: Parton distributions; QCD evolution equation; Nonlinear
corrections

\newpage
\begin{center}
\section{Introduction}
\end{center}

The parton distributions of nucleons are important components of
our understanding of high energy physics. The $Q^2$ dependence of parton densities was predicted
based on a rigorous application of the renormalization
group equations in QCD, guaranteeing the factorization of parton-densities
in \cite{1,2,3}. Currently the available parton distributions were extracted from the
experimental data with the linear QCD evolution
equation-the Dokshitzer-Gribov-Lipatov-Altarelli-Parisi (DGLAP)
equation \cite{1,2,3,4,5,6}, which describes all results of \cite{1,2,3} using
an illuminating partonic picture.
The solutions of the DGLAP equations depend on
the initial parton distributions at low starting scale $\mu^2$.
There are two different choices for the input distributions: (1) in
global analysis, the starting
point is fixed at an arbitrary scale $Q_0^2
> 1~ $GeV$^2$ and the corresponding input parton distributions are
parameterized by comparing with the measured data at $Q^2>Q_0^2$.
These input distributions are irrelevant to any physics models,
one may even input negative gluon distributions; (2) in dynamical
models, the parton distributions at $Q^2 >1~$GeV$^2$ are generated by QCD
radiative corrections generated from imaginary intrinsic parton
distributions at an optimally determined $Q^2_0< 1~$GeV$^2$ according
to the nucleon model. For example, it is well known that the nucleons
consist of three constituents at very low $Q^2$. A
natural attempt, firstly proposed in 1977, is to assume that
the nucleons consist of valence quarks at low starting
point $\mu^2$ (but still in the perturbative region
$\alpha_s(\mu^2)/2\pi <1$ and $\mu>\Lambda_{QCD}$), and the gluons
and sea quarks are radioactively produced at $Q^2>\mu^2$ \cite{7,8,9}. These
input distributions allow one to construct a complete QCD picture of the proton \cite{10,11,12,13,14,15}.
However, such natural inputs fail due to overly steep behavior of the
predicted parton distributions at the small Bjorken variable $x$. Instead of the natural
inputs, Reya, Gl\"{u}ck and Vogt (GRV) \cite{16} added the valance
like sea quarks and gluon distributions to the input parton
distributions at a little larger $Q^2$ scale. The predictions of the
GRV model are compatible with the data at $Q^2>4~ $GeV$^2$ and
$x>10^{-4}$.

    All the above approaches determine the parton distributions via the DGLAP
equation. Comparing with the natural input distributions, the
valence-like distributions of the gluon and sea quarks in the
GRV-model can slow down the evolution of the DGLAP equation at low
$Q^2$ and reach agreement with experimental results, since the evolution region
of the valence like distributions is sizeably larger. On the other
hand, we know that the contributions from the parton recombination
corrections become important at $Q^2<1~ $GeV$^2$, which are
neglected in the DGLAP equation. The correlation among initial
partons can not be neglected towards small $x$ and low $Q^2$. The
negative corrections of the parton recombination also slow down the
partons evolution. These nonlinear effects can be calculated in the
perturbative QCD. The nonlinear corrections of the gluon
recombination to the DGLAP equation were firstly derived by Gribov,
Levin and Ryskin \cite{17} and by Mueller and Qiu \cite{18} in the
double leading logarithmic (DLL) approximation (see Appendix A).
The similar research for the nonlinear corrections to the DGLAP equation was
discussed by many authors \cite{19,20,21}. In particularly,
this evolution
equation was re-derived to include parton recombination at all
$x$ by Zhu, Ruan and Shen \cite{22,23,24} in the leading
logarithmic (LL(Q$^2$)) approximation. We refer to this version
of the nonlinear corrections as the ZRS corrections to the
DGLAP equation. Although the relation between the derivation of the nonlinear
part and the renormalization
group theory is unclear, in this approach \cite{22,23,24} all stages of the calculation refer to parton concepts
and offer a very illuminating physical interpretation of the parton recombination using the time ordered perturbative theory (TOPT).

    The success of the GRV model inspires us to use natural inputs to
replace the valence like inputs in the GRV model due to
the nonlinear corrections of the parton recombination. Our main
results are: (i) we find that the ZRS corrections suppress the
fast increase of the sea quark- and gluon-densities using the
natural input and has similar results as the GRV(98LO) \cite{25}  at
$x>10^{-4}$ and $Q^2> 4 ~$GeV$^2$; (ii) we predict that the parton
distributions at $Q^2<1 ~$GeV$^2$ are positively defined,
particularly, the sea quark distributions appear a plateau at small
$x$ and low $Q^2$, indicating Pomeron-like behavior \cite{26}; (iii)
our input quark distributions are compatible with the valence quark
distributions predicted by effective chiral quark model \cite{27};
(iv) this evidence of the parton recombination existing in the
standard QCD evolution provides a possible dynamical way to explore
the nuclear shadowing effects.

    The organization of this paper is as follows. We present
the ZRS corrections to the DGLAP equation in Sec. 2. Using
the nonlinear QCD evolution equation and the natural input
distributions, we calculate the evolution of the parton
distributions in the proton, and the corresponding parameters are
discussed in Sec. 3. Our resulting parton distributions comparing
with the experimental data and some databases are presented
in Sec. 4. Section 5 is the discussion and summary, where the
applicability of the DGLAP equation with the ZRS corrections at the low $Q^2$
is discussed.

\newpage
\begin{center}
\section{The nonlinear QCD evolution equation}
\end{center}

   The DGLAP equation predicts a strong rise in the parton
densities when the Bjorken variable $x$ decrease toward small values due to
the elementary process is one-parton splitting to two-partons.
This behavior violates unitarity. One can expect that at very large number densities of partons, for example in the small $x$ region, the
wave functions of partons can overlap. In this case the contributions of
two-partons-to-two-partons subprocesses (i.e.,  parton recombination) should be considered in the
QCD evolution equations. Various models are proposed to modify the twist-2
DGLAP evolution kernels (i.e., the parton splitting functions).
The derivation of such equations needs to sum the
contributions from real and interference Feynman diagrams and corresponding virtual
diagrams. Gribov, Levin and Ryskin in \cite{17} use the AGK cutting rule \cite{28} to count the
contributions of interference diagrams. Later Mueller and
Qiu \cite{18} calculate the (real) gluon recombination functions at the double leading logarithmic
approximation (DLLA) in a covariant perturbation framework.

\begin{figure}[htp]
\centering
\includegraphics[width=0.5\textwidth]{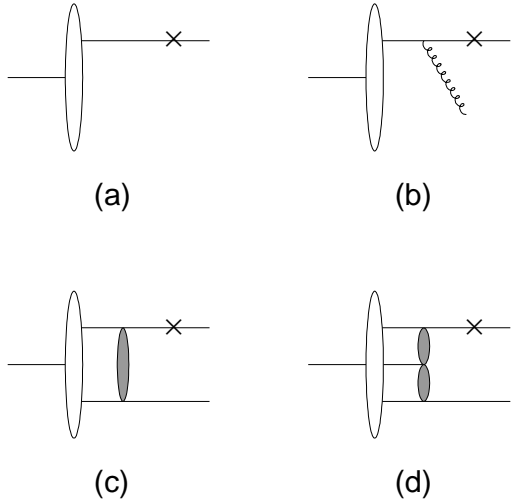}
\caption{The elemental amplitudes, which contribute to the ZRS
corrections, i.e., the corrections of the parton recombination to the
standard DGLAP equation; where we have omitted the distinction
between the parton flavors, "x" means the probing place and the dark
circles indicate QCD interactions among all the possible correlating
partons.}
\label{fig1}
\end{figure}

    However, the following motivations led Zhu and his cooperators to re-derive the above
QCD evolution equation with the parton recombination corrections:
(1) the application of the AGK cutting rule in the GLR-MQ corrections breaks the evolution kernels (see Fig. 1 in Ref. \cite{22});
(2) the GLR-MQ corrections to the DGLAP equation violate the momentum conservation\cite{29};
(3) the DLL approximation is valid only at small $x$ and the GLR-MQ corrections can not smoothly connect with the DGLAP equation \cite{22}.

    To avoid these disadvantages, the
relations among the relevant high twist-amplitudes are derived by using the TOPT approach at the LL($Q^2$) approximation
in a same framework as the derivation of the DGLAP equation in \cite{22}, where the TOPT cutting rules are proposed
to connect with various cut diagrams instead of the AGK cutting rules. Thus, one can obtain the complete
contributions only using the calculations of the real cut diagrams. As a consequence, a set of new evolution
equations with parton recombination in a general framework was established. In next step, the
the recombination functions are calculated in a whole $x$ region based on the same
TOPT-framework \cite{23,24}.

      According to the above mentioned statement, we arrange
the QCD evolution equation including the ZRS corrections at
the LL(Q$^2$) approximation for the parton distributions in a proton
as follows.

       We denote $f_{v_j}(x,Q^2)$ (j=u,d) for valence quark distributions, $f_{q_i}(x,Q^2)$ (i=u,d,s)
for sea quark distributions, $f_{\overline{q}_i}(x,Q^2)$ (i=u,d,s)
for anti-sea quark distributions and $f_g(x,Q^2)$ for gluon
distribution. We define $\Sigma(x,Q^2)\equiv\sum_j
f_{v_j}(x,Q^2)+\sum_i
f_{q_i}(x,Q^2)+\sum_if_{\overline{q}_i}(x,Q^2)$.
We start from the
elemental amplitudes Fig. 1. According to Ref. \cite{22}, the
DGLAP equation with the ZRS corrections reads

$$Q^2\frac{dxf_{v_j}(x,Q^2)}{dQ^2}$$
$$=\frac{\alpha_s(Q^2)}{2\pi}\int_x^1\frac{dy}{y}P_{qq}(z)xf_{v_j}(y,
Q^2)$$
$$-\frac{\alpha_s(Q^2)}{2\pi}xf_{v_j}(x,Q^2)\int_0^1 dzP_{qq}(z)$$
$$-\frac{\alpha_s^2(Q^2)}{4\pi R^2 Q^2}\int_x^{1/2}\frac{dy}{y}xP_{qg\rightarrow q}(x,y)yf_g(y,Q^2)yf_{v_j}(y,Q^2)$$
$$+\frac{\alpha_s^2(Q^2)}{4\pi R^2 Q^2}\int_{x/2}^x\frac{dy}{y}xP_{qg\rightarrow q}(x,y)yf_g(y,Q^2)yf_{v_j}(y,Q^2)$$
$$-\frac{\alpha_s^2(Q^2)}{4\pi R^2 Q^2}\int_x^{1/2}\frac{dy}{y}xP_{qq\rightarrow q}(x,y)y[\Sigma(y,Q^2)-f_{v_j}(y,Q^2)]yf_{v_j}(y,Q^2)$$
$$+\frac{\alpha_s^2(Q^2)}{4\pi R^2 Q^2}\int_{x/2}^x\frac{dy}{y}xP_{qq\rightarrow q}(x,y)y[\Sigma(y,Q^2)-f_{v_j}(y,Q^2)]yf_{v_j}(y,Q^2),(if~x\le 1/2), $$

$$Q^2\frac{dxf_{v_j}(x,Q^2)}{dQ^2}$$
$$=\frac{\alpha_s(Q^2)}{2\pi}\int_x^1\frac{dy}{y}P_{qq}(z)xf_{v_j}(y,
Q^2)$$
$$-\frac{\alpha_s(Q^2)}{2\pi}xf_{v_j}(x,Q^2)\int_0^1 dzP_{qq}(z)$$
$$+\frac{\alpha_s^2(Q^2)}{4\pi R^2 Q^2}\int_{x/2}^{1/2}\frac{dy}{y}xP_{qg\rightarrow q}(x,y)yf_g(y,Q^2)yf_{v_j}(y,Q^2)$$
$$+\frac{\alpha_s^2(Q^2)}{4\pi R^2 Q^2}\int_{x/2}^{1/2}\frac{dy}{y}xP_{qq\rightarrow q}(x,y)y[\Sigma(y,Q^2)-f_{v_j}(y,Q^2)]yf_{v_j}(y,Q^2),(if~1/2\le x\le 1), \eqno(1-a)$$
for valence quarks, where $z=x/y$, the factor $1/(4\pi R^2)$ is from normalizing two-parton distribution,
R is the correlation length of two initial partons.

$$Q^2\frac{dxf_{\overline{q}_i}(x,Q^2)}{dQ^2}$$
$$=\frac{\alpha_s(Q^2)}{2\pi}\int_x^1\frac{dy}{y}P_{qq}(z)xf_{\overline{q}_i}(y,
Q^2)$$
$$-\frac{\alpha_s(Q^2)}{2\pi}xf_{\overline{q}_i}(x,Q^2)\int_0^1dzP_{qq}(z)$$
$$+\frac{\alpha_s(Q^2)}{2\pi}\int_x^1\frac{dy}{y}P_{qg}(z)xf_g(y,
Q^2)$$
$$-\frac{\alpha_s^2(Q^2)}{4\pi R^2 Q^2}\int_x^{1/2}\frac{dy}{y}x
P_{gg\rightarrow \overline{q}}(x,y)[ yf_g(y,Q^2)]^2$$
$$+\frac{\alpha_s^2(Q^2)}{4\pi R^2 Q^2}\int_{x/2}^x\frac{dy}{y}x
P_{gg\rightarrow \overline{q}}(x,y)[ yf_g(y,Q^2)]^2$$
$$-\frac{\alpha_s^2(Q^2)}{4\pi R^2 Q^2}\int_x^{1/2}\frac{dy}{y}x
P_{q\overline{q}\rightarrow \overline{q}}(x,y)yf_{q_i}(y,Q^2)yf_{\overline{q}_i}(y,Q^2)$$
$$+\frac{\alpha_s^2(Q^2)}{4\pi R^2 Q^2}\int_{x/2}^{x}\frac{dy}{y}x
P_{q\overline{q}\rightarrow \overline{q}}(x,y)yf_{q_i}(y,Q^2)yf_{\overline{q}_i}(y,Q^2)$$
$$-\frac{\alpha_s^2(Q^2)}{4\pi R^2 Q^2}\int_x^{1/2}\frac{dy}{y}x
P_{\overline{q} \overline{q}\rightarrow \overline{q}}(x,y)y[\Sigma(y,Q^2)-f_{q_i}(y,Q^2)]yf_{\overline{q}_i}(y,Q^2)$$
$$+\frac{\alpha_s^2(Q^2)}{4\pi R^2 Q^2}\int_{x/2}^{x}\frac{dy}{y}x
P_{\overline{q} \overline{q}\rightarrow\overline{q}}(x,y)y[\Sigma(y,Q^2)-f_{q_i}(y,Q^2)]yf_{\overline{q}_i}(y,Q^2)$$
$$-\frac{\alpha_s^2(Q^2)}{4\pi R^2 Q^2}\int_x^{1/2}\frac{dy}{y}x
P_{\overline{q}g\rightarrow \overline{q}}(x,y)yf_g(y,Q^2)yf_{\overline{q}_i}(y,Q^2)$$
$$+\frac{\alpha_s^2(Q^2)}{4\pi R^2 Q^2}\int_{x/2}^x\frac{dy}{y}x
P_{\overline{q}g\rightarrow \overline{q}}(x,y) yf_g(y,Q^2)yf_{\overline{q}_i}(y,Q^2),(if~x\le 1/2),$$

$$Q^2\frac{dxf_{\overline{q}_i}(x,Q^2)}{dQ^2}$$
$$=\frac{\alpha_s(Q^2)}{2\pi}\int_x^1\frac{dy}{y}P_{qq}(z)xf_{\overline{q}_i}(y,
Q^2)$$
$$-\frac{\alpha_s(Q^2)}{2\pi}xf_{\overline{q}_i}(x,Q^2)\int_0^1dzP_{qq}(z)$$
$$+\frac{\alpha_s(Q^2)}{2\pi}\int_x^1\frac{dy}{y}P_{qg}(z)xf_g(y,
Q^2)$$
$$+\frac{\alpha_s^2(Q^2)}{4\pi R^2 Q^2}\int_{x/2}^{1/2}\frac{dy}{y}x
P_{gg\rightarrow \overline{q}}(x,y)[ yf_g(y,Q^2)]^2$$
$$+\frac{\alpha_s^2(Q^2)}{4\pi R^2 Q^2}\int_{x/2}^{1/2}\frac{dy}{y}x
P_{q\overline{q}\rightarrow \overline{q}}(x,y)yf_{q_i}(y,Q^2)yf_{\overline{q}_i}(y,Q^2)$$
$$+\frac{\alpha_s^2(Q^2)}{4\pi R^2 Q^2}\int_{x/2}^{1/2}\frac{dy}{y}x
P_{\overline{q} \overline{q}\rightarrow \overline{q}}(x,y)y[\Sigma(y,Q^2)-f_{q_i}(y,Q^2)]yf_{\overline{q}_i}(y,Q^2)$$
$$+\frac{\alpha_s^2(Q^2)}{4\pi R^2 Q^2}\int_{x/2}^{1/2}\frac{dy}{y}x
P_{\overline{q}g\rightarrow \overline{q}}(x,y)yf_g(y,Q^2)yf_{\overline{q}_i}(y,Q^2),(if~1/2\le x\le 1),\eqno(1-b)$$
for sea quark distributions and

$$Q^2\frac{dxf_g(x,Q^2)}{dQ^2}$$
$$=\frac{\alpha_s(Q^2)}{2\pi}\int_x^1\frac{dy}{y}P_{gq}(z)x\Sigma(y,
Q^2)$$
$$+\frac{\alpha_s(Q^2)}{2\pi}\int_x^1\frac{dy}{y}P_{gg}(z)xf_g(y,
Q^2)$$
$$-f\frac{\alpha_s(Q^2)}{2\pi}xf_g(x,Q^2)\int_0^1dzP_{qg}(z)$$
$$-\frac{1}{2}\frac{\alpha_s(Q^2)}{2\pi}xf_g(x,Q^2)\int_0^1 dzP_{gg}(z)$$
$$-\frac{\alpha_s^2(Q^2)}{4\pi R^2 Q^2}\int_x^{1/2}\frac{dy}{y}x
P_{gg\rightarrow g}(x,y)[ yf_g(y,Q^2)]^2$$
$$+\frac{\alpha_s^2(Q^2)}{4\pi R^2 Q^2}\int_{x/2}^x\frac{dy}{y}x
P_{gg\rightarrow g}(x,y)[ yf_g(y,Q^2)]^2$$
$$-\frac{\alpha_s^2(Q^2)}{4\pi R^2 Q^2}\int_x^{1/2}\frac{dy}{y}x
P_{q\overline{q}\rightarrow g}(x,y)\sum_{i=1}^{f}[ yf_{\overline{q}_i}(y,Q^2)]^2$$
$$+\frac{\alpha_s^2(Q^2)}{4\pi R^2 Q^2}\int_{x/2}^x\frac{dy}{y}x
P_{q\overline{q}\rightarrow g}(x,y)\sum_{i=1}^{f}[ yf_{\overline{q}_i}(y,Q^2)]^2$$
$$-\frac{\alpha_s^2(Q^2)}{4\pi R^2 Q^2}\int_x^{1/2}\frac{dy}{y}x
P_{qg\rightarrow g}(x,y)y\Sigma(y,Q^2)yf_g(y,Q^2)$$
$$+\frac{\alpha_s^2(Q^2)}{4\pi R^2 Q^2}\int_{x/2}^x\frac{dy}{y}x
P_{qg\rightarrow g}(x,y)y\Sigma(y,Q^2)yf_g(y,Q^2),(if~x\le 1/2),$$

$$Q^2\frac{dxf_g(x,Q^2)}{dQ^2}$$
$$=\frac{\alpha_s(Q^2)}{2\pi}\int_x^1\frac{dy}{y}P_{gq}(z)x\Sigma(y,
Q^2)$$
$$+\frac{\alpha_s(Q^2)}{2\pi}\int_x^1\frac{dy}{y}P_{gg}(z)xf_g(y,
Q^2)$$
$$-f\frac{\alpha_s(Q^2)}{2\pi}xf_g(x,Q^2)\int_0^1dzP_{qg}(z)$$
$$-\frac{1}{2}\frac{\alpha_s(Q^2)}{2\pi}xf_g(x,Q^2)\int_0^1 dzP_{gg}(z)$$
$$+\frac{\alpha_s^2(Q^2)}{4\pi R^2 Q^2}\int_{x/2}^{1/2}\frac{dy}{y}x
P_{gg\rightarrow g}(x,y)[ yf_g(y,Q^2)]^2$$
$$+\frac{\alpha_s^2(Q^2)}{4\pi R^2 Q^2}\int_{x/2}^{1/2}\frac{dy}{y}x
P_{q\overline{q}\rightarrow g}(x,y)\sum_{i=1}^{f}[ yf_{\overline{q}_i}(y,Q^2)]^2$$
$$+\frac{\alpha_s^2(Q^2)}{4\pi R^2 Q^2}\int_{x/2}^{1/2}\frac{dy}{y}x
P_{qg\rightarrow g}(x,y)y\Sigma(y,Q^2)yf_g(y,Q^2),
(if~1/2\le x\le 1),\eqno(1-c)$$ for gluon distribution. The corresponding cut diagrams are presented in Fig. 2. Note that the TOPT cutting rules \cite{22} are used in Eq. (1). Thus, we need only compute the two-partons to two-partons kernels and use the same kernels to write the contributions of the interference processes. On the other hand, the contributions of the virtual processes are canceled each other \cite{22}.

\begin{figure}[htp]
\centering
\includegraphics[width=0.55\textwidth]{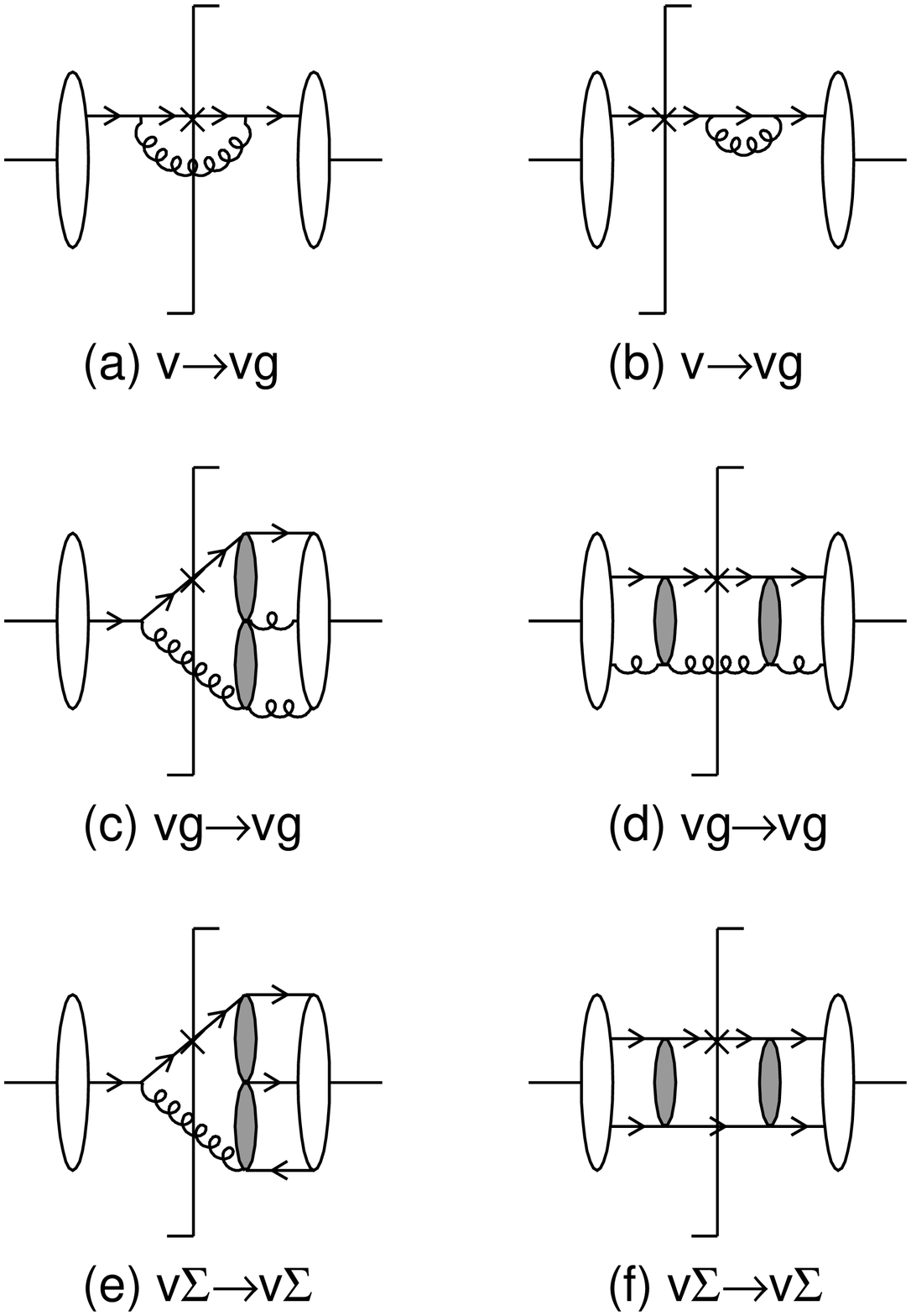}
\includegraphics[width=0.6\textwidth]{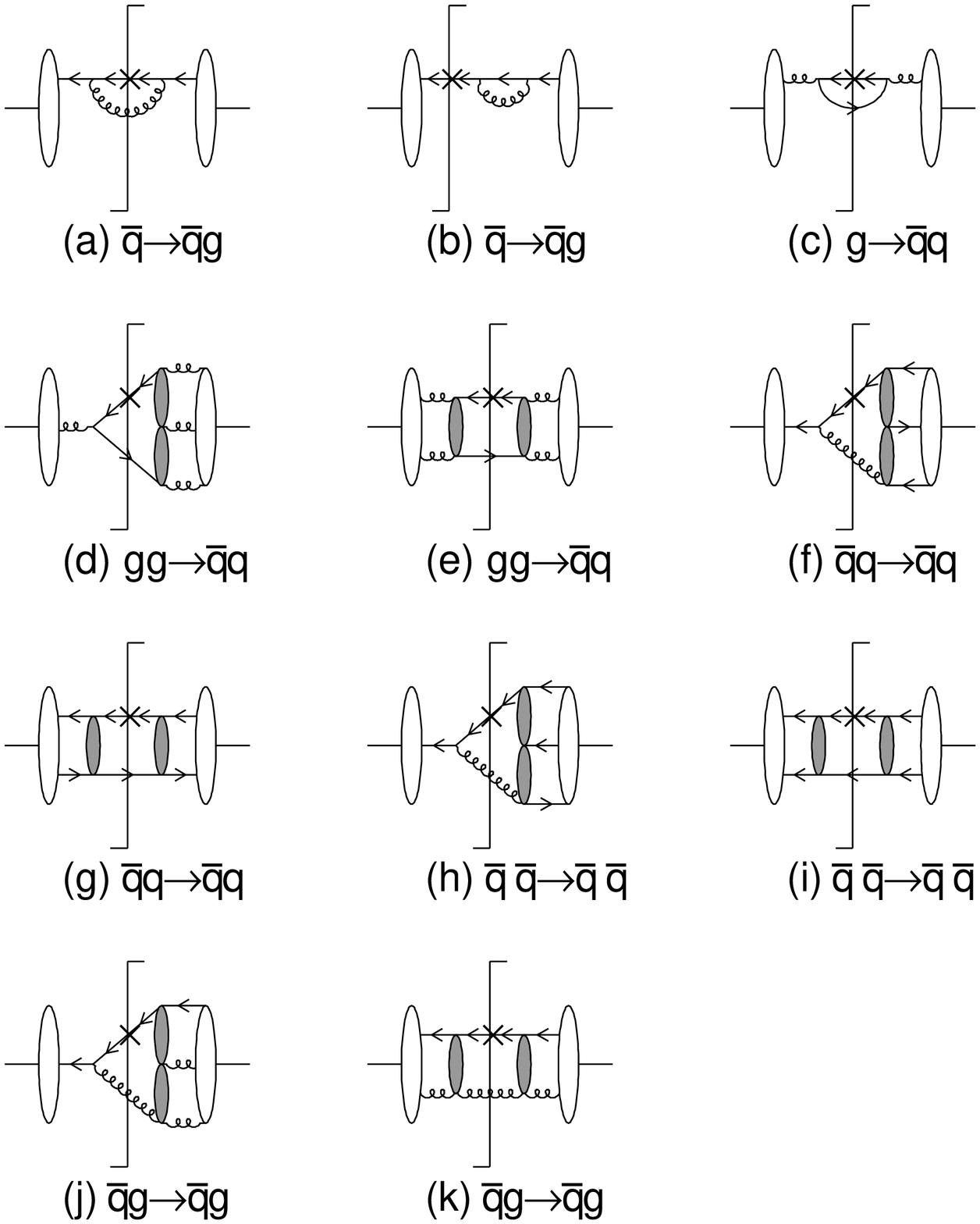}
\label{fig2}
\end{figure}

\begin{figure}[htp]
\centering
\includegraphics[width=0.55\textwidth]{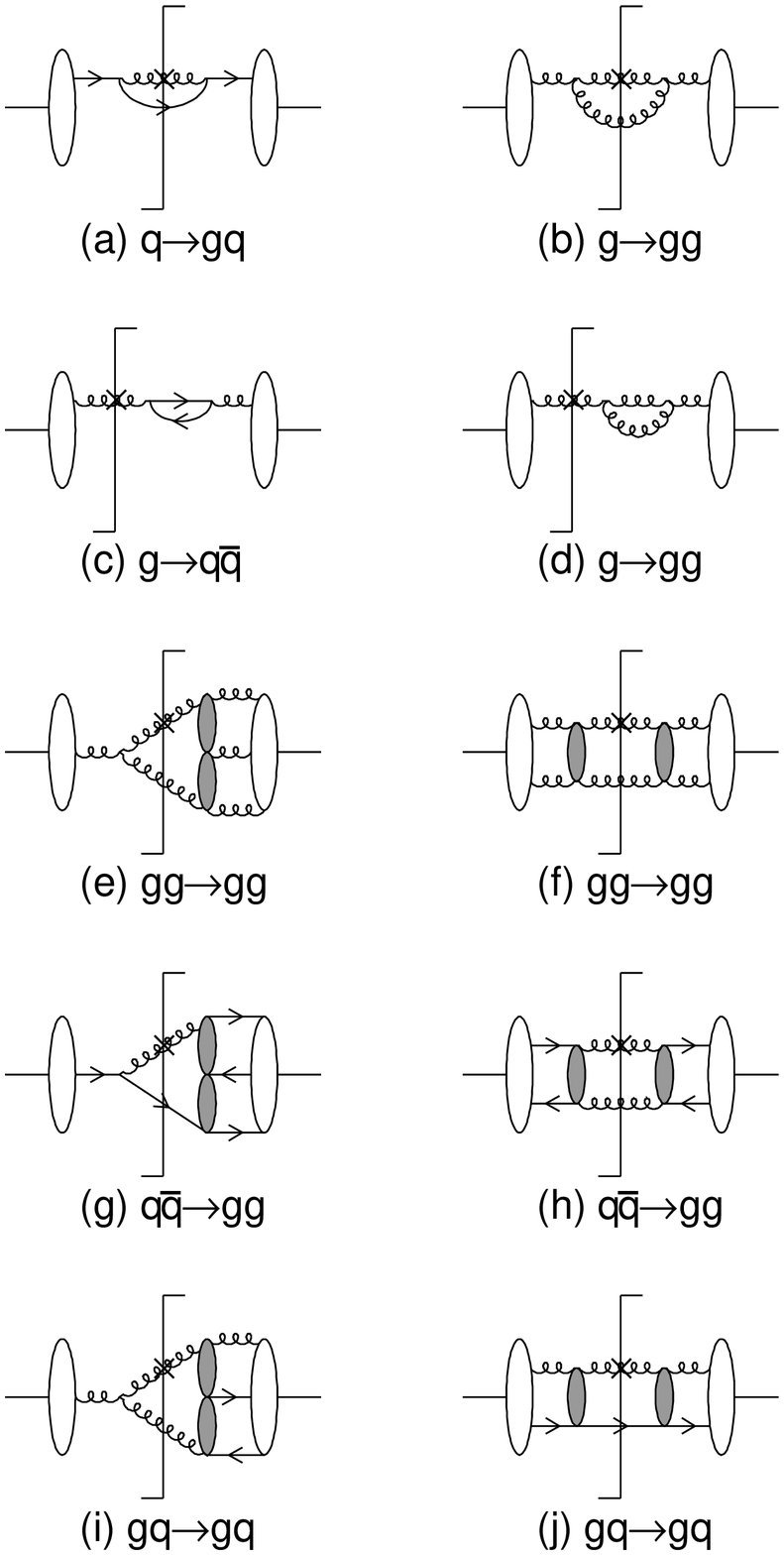}
\caption{The cut diagrams for the DGLAP equation with the ZRS corrections using the
elemental amplitudes in Fig. 1.}
\label{fig2}
\end{figure}

    The un-regularized DGLAP splitting kernels in the linear terms are \cite{22,30,31}

$$P_{gg}(z)=2C_A\left[z(1-z)+\frac{1-z}{z}+\frac{z}{1-z}\right],$$

$$P_{gq}(z)=C_F\frac{1+(1-z)^2}{z},$$

$$P_{qq}(z)=C_F\frac{1+z^2}{1-z},$$

$$P_{qg}(z)=T_R[z^2+(1-z)^2],\eqno(2)$$ where
$C_A=N_c=3,~~C_F=\frac{N_c^2-1}{2N_c}=\frac{4}{3},~~T_R=\frac{1}{2}$ and $z=x/y$.

    The recombination functions (Fig. 3) in the nonlinear terms are

$$P_{gg\rightarrow
g}(x,y)=\frac{9}{64}\frac{(2y-x)(72y^4-48xy^3+140x^2y^2-116x^3y+29x^4)}{xy^5},$$

$$P_{gg\rightarrow q}(x,y)=P_{gg\rightarrow \overline{q}}(x,y)=\frac{1}{96}\frac{(2y-x)^2(18y^2-21xy+14x^2)}{y^5},$$

$$P_{qq\rightarrow q}(x,y)=P_{\overline{q}\overline{q}\rightarrow\overline{q}}(x,y)=
\frac{2}{9}\frac{(2y-x)^2}{y^3},$$

$$P_{q\overline{q}\rightarrow q}(x,y)=P_{q\overline{q}\rightarrow \overline{q}}(x,y)=
\frac{1}{108}\frac{(2y-x)^2(6y^2+xy+3x^2)}{y^5},$$

$$P_{qg\rightarrow
q}(x,y)=P_{\overline{q}g\rightarrow\overline{q}}(x,y)=\frac{1}{288}\frac{(2y-x)(140y^2-52yx+65x^2)}{y^4},$$

$$P_{qg\rightarrow
g}(x,y)=P_{\overline{q}g\rightarrow g}(x,y)=\frac{1}{288}\frac{(2y-x)(304y^2-202yx+79x^2)}{xy^4},$$

$$P_{q\overline{q}\rightarrow
g}(x,y)=\frac{4}{27}\frac{(2y-x)(18y^2-9yx+4x^2)}{xy^3}, \eqno(3)$$
they are taken from Ref. \cite{23}.

\begin{figure}[htp]
\centering
\includegraphics[width=0.6\textwidth]{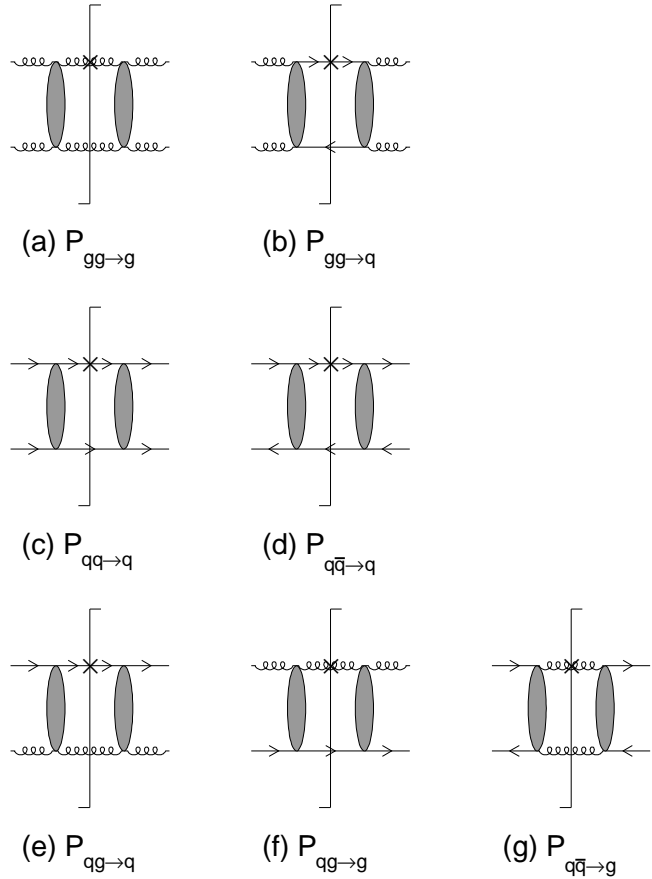}
\caption{The recombination functions in Eq. (3), which are calculated at LL($Q^2$)A in Ref.\cite{12}.}
\label{fig3}
\end{figure}

    The properties of the equation (1-3) are summarized as follows.

    (1) The QCD evolution kernels
(for example, splitting function in the DGLAP equation and
recombination functions in the corrected equation) are
separated from the coefficient functions using the equivalent
particle approximation\cite{32,33,34,35}. However, the infrared (IR)
divergence due to the gauge singularities in some twist-4 amplitudes
prevents us from using the equivalent particle approximation, since
these gauge terms are coupled with the backward components of two
parton legs which connect with the probe. In the derivation of the
GLR-MQ corrections, the gauge singularities become finite using a
contour integral. In this method, the backward components of two
legs still exist, i.e., the propagators of the legs are
off-mass-shell. This implies that the equivalent particle
approximation is invalid. On the other hand, the gauge singularities
in the twist-4 coefficient functions are safely removed from the
ZRS corrections using the TOPT method. Thus, the recombination
kernels can be simply separated from the coefficient functions at
the equivalent particle approximation. A detailed discussion see Ref.
\cite{24}.

    (2) The equation is a result summing all possible cut diagrams up to twist-4
in a quantum field theory framework (i.e., the TOPT) instead of using the AGK cutting rules.
The two-parton-to-two-parton amplitude leads to the positive (antishadowing)
effect, while the contributions of interference amplitudes between
the one-parton-to-two-partons and the three-partons-to-two-partons processes yield a
negative (shadowing) effect.
The coexistence of shadowing and antishadowing in the QCD evolution
of the parton densities is a general requisition of the
local momentum conservation \cite{29}. We emphasize that the shadowing
and antishadowing terms are defined on different kinematic
domains $[x,1/2]$ and $[x/2,x]$, respectively. Thus, the momentum is conservative
as shown in following Eq. (4). On the other hand, the AGK cutting rule
is used in the derivation of the GLR-MQ corrections, where the contributions of
the positive and negative terms only differ
in the numerical weights. Thus, the antishadowing effect in the GLR-MQ corrections is completely canceled by the
effect and the resulting evolution equation violates the momentum conservation.

    (3) Since Eqs. (1)-(3) are derived at the LL$(Q^2)$ approximation and they contain the terms beyond the
DLL approximation, the DGLAP equation with the ZRS corrections is valid in the full-$x$
range if we neglect the Balitsky- Fadin-Kuraev-Lipatov (BFKL)
corrections \cite{36,37,38,39,40,41} at the very small-$x$ region. Thus, the
ZRS corrections can smoothly connect with the DGLAP equation.

    (4) The sea quark evolutions in the ZRS corrections (1-b) and GLR-MQ corrections (A-2) (see Appendix A)
take different forms. The reason is that the transition of
gluon-quarks is suppressed in the DLLA manner. The DLLA diagram
contains only the gluon ladders and any transitions of gluon-quark
break the gluon ladder structure. Therefore, a special box diagram
is used to include the corrections of gluon recombination to the
quark distributions in the GLR-MQ corrections. However, this extra
diagram is unnecessary in the derivation of the ZRS corrections,
since we can produce the evolution equations for gluon and sea
quarks in the same framework at the LL$(Q^2)$ approximation.

    (5) The combining distribution of two
partons, for say two gluons, is assumed to be $f_g^{(2)}(y,Q^2)\sim
f_g^2(y,Q^2)$ either in the ZRS corrections or in the GLR-MQ
corrections. This is a simplest model. One of us (W.Z.) has discussed
the recombination of gluons with different values of $x$ in the
nonlinear evolution equation and finds that this modification
unreasonably enhances the shadowing effect in the GLR-MQ corrections,
while it does not change the predictions of the ZRS corrections,
since the momentum conservation plays an important role in this
result \cite{42}.

     Comparing with the GLR-MQ corrections, the ZRS corrections reasonably describe the corrections
of the parton recombination to the DGLAP equation at the twist-4
level. The DGLAP equation with the ZRS corrections has been applied to study some small
$x$ phenomena in its approximation form \cite{43,44,45,46}. In this work we
will use its complete version Eq. (1-3) to quantitatively study the
parton distributions in the proton.

\newpage
\begin{center}
\section{Natural initial distributions and free parameters fitting }
\end{center}

    The solutions of the QCD evolution equations for the parton distributions
depend on the initial parton distributions at a low scale $Q^2=\mu^2$. An ideal and
simple assumption is that the nucleon consists entirely of
three valence quarks at $\mu^2$ and the gluon and sea distributions
at $Q^2>\mu^2$ are generated radioactively. This naive model was
firstly proposed in Refs. \cite{7,8,9} and it successfully predicts that at
low $Q^2\sim 1~ $GeV$^2$ about $50\%$ of the nucleon
momentum is already carried by gluons, which
agrees with the experimental results. Unfortunately, the
distributions of the sea quarks and gluon predicted by this natural
input and the DGLAP equation are too steep. Instead of
the simple input distributions, Gl\"{u}ck, Reya and Vogt
(GRV) \cite{16} assumed that the nucleon has valence quarks and special
valence-like sea quarks and valence-like gluon at a starting point
at $Q^2\simeq 0.2\sim 0.3~ $GeV$^2$. The GRV input in the DGLAP
equation nicely predicts the parton distributions in a broad
kinematical range.

    Now we go back to the natural input distributions with the ZRS corrections
using Eqs. (1-3) for the second moments of the distributions and the
measured momentum of the valence quark distributions at a higher
$Q^2$, we obtain the starting point $\mu^2=0.064 ~$GeV$^2$ (with
$\Lambda_{QCD} =0.204 ~$GeV for f=3 flavors). Note that the value of
$1/\mu\sim 0.8 fm$ is compatible with a typical radius of the proton.  This value of $\mu$ is similar
to the previous estimation in Refs. \cite{7,8,9}, where the DGLAP
equation is used. The reason is that the contributions of the
nonlinear (shadowing and antishadowing) terms to the $Q^2$-evolution
of the momentum fractions in the ZRS corrections vanish due to
the momentum conservation \cite{22}

 $$-\frac{\alpha_s^2(Q^2)}{4\pi R^2Q^2}\int_0^{1/2}dx\int_x^{1/2}\frac{dy}{y}xP_{qg\rightarrow q}(z)[ yf_g(y,Q^2)yf_v(y,Q^2)]$$
$$+\frac{\alpha_s^2(Q^2)}{4\pi R^2Q^2}\int_0^{1/2}dx\int_{x/2}^x\frac{dy}{y}xP_{qg\rightarrow q}(z)[ yf_g(y,Q^2)yf_v(y,Q^2)]$$
$$+\frac{\alpha_s^2(Q^2)}{4\pi R^2Q^2}\int_{1/2}^1dx\int_{x/2}^{1/2}\frac{dy}{y}xP_{qg\rightarrow q}(z)[ yf_g(y,Q^2)yf_v(y,Q^2)]=0. \eqno(4)$$

    According to the natural input distribution,

$$f_g(x,\mu^2)=0,~~f_{q_i}(x,\mu^2)=f_{\overline {q}_i}(x,\mu^2)=0. \eqno(5)$$
We choose a minimum free parameter scheme for the typical valence quark distributions

$$
xf_{v_u}(x,\mu^2)=A_ux^{B_u}(1-x)^{C_u},
$$

$$
xf_{v_d}(x,\mu^2)=A_dx^{B_d}(1-x)^{C_d}, \eqno(6)$$ which
satisfy the momentum sum rule

$$\int_0^1dxx[f_{v_u}(x,\mu^2)+f_{v_d}(x,\mu^2)]=1. \eqno(7)$$ and the normalization conditions
$$\int_0^1dxf_{v_u}(x,\mu^2)=2,~~\int_0^1dxf_{v_d}(x,\mu^2)=1.\eqno(8)$$
Fitting to experimental data,
we obtain following values for the proton
$A_u=24.30,~B_u=1.98,~~ C_u=2.06,$~~$A_d=9.10,~~B_d=1.31$, and
$C_d=3.80$.

\begin{figure}[htp]
\centering
\includegraphics[width=0.5\textwidth]{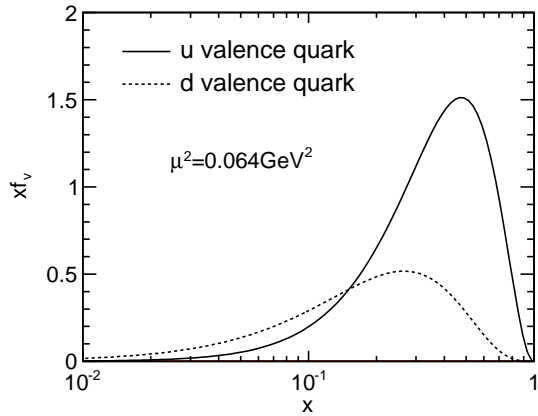}
\caption{The natural input distributions in the proton at the initial
scale $\mu^2 =0.064 $ GeV$^2$.}
\label{fig4}
\end{figure}

    The corresponding input distributions are shown in Fig. 4. The
free parameter $R$ in Eq.(1) depends on the geometric
distributions of partons inside the proton. $R\simeq 5~ $GeV$^{-1}$
(or $R< 5 ~$GeV$^{-1}$) when the partons distribute uniformly (or non-uniformly)
in the proton. We take $R=4.24~ $GeV$^{-1}$ for best fitting to the data.

\newpage
\begin{center}
\section{Parton distributions in the proton}
\end{center}

    Since the four free parameters are fixed, now it is straightforward
to calculate the evolution of the parton
distributions in the proton using Eqs. (1-3). The $x$ and $Q^2$
dependence of our predicted structure functions and comparisons with
the data \cite{47,48,49,50,51,52,53,54,55,56,57,58,59} are shown in Figs. 5 and 6.
Our results are generally lower than the experimental data at $x<10^{-4}$. It
implies that the BFKL correction
is not negligible in such small $x$ range.

\begin{figure}[htp]
\centering
\includegraphics[width=0.6\textwidth]{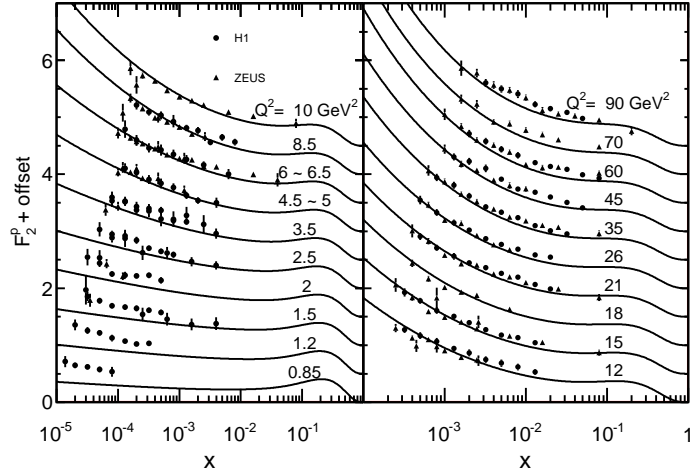}
\caption{Comparisons of our predicted $x$-dependence of $F_2^p(x,Q^2)$
(solid curves) with HERA data \cite{47,48,49,50,51,52,53,54,55,56} at small $x$.}
\label{fig5}
\end{figure}

\begin{figure}[htp]
\centering
\includegraphics[width=0.6\textwidth]{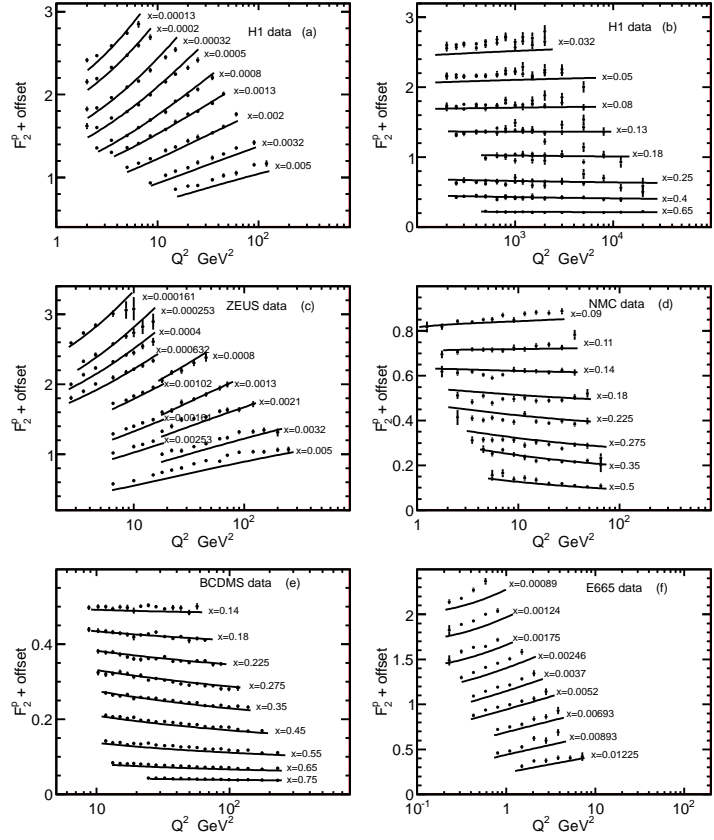}
\caption{Comparisons of our predicted $Q^2$-dependence of
$F_2^p(x,Q^2)$ (solid curves) with various experimental data
\cite{47,48,49,50,51,52,53,54,55,56,57,58,59}.}
\label{fig6}
\end{figure}

\begin{figure}[htp]
\centering
\includegraphics[width=0.6\textwidth]{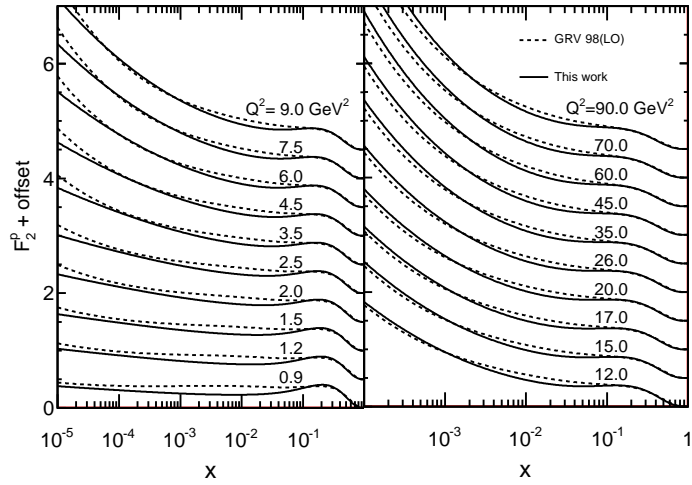}
\caption{Comparisons of the $x$-dependence of our predicted $F_2^P$ (solid curves)
with the GRV(98LO) results (dashed curves) \cite{25}.}
\label{fig7}
\end{figure}

    The comparisons of two dynamically generated parton distributions of the proton
(i.e., using Eqs. (1-3) with the natural input and
using the DGLAP equation with the GRV(98LO) input) are shown in
Figs. 7-10. In Figs. 9 and 10 we plot the parton distributions in
the linear DGLAP equation with the natural input. One can find that
the valence like input distributions in the GRV model are equivalent
to the effective description of the nonlinear parton recombination.

\begin{figure}[htp]
\centering
\includegraphics[width=0.7\textwidth]{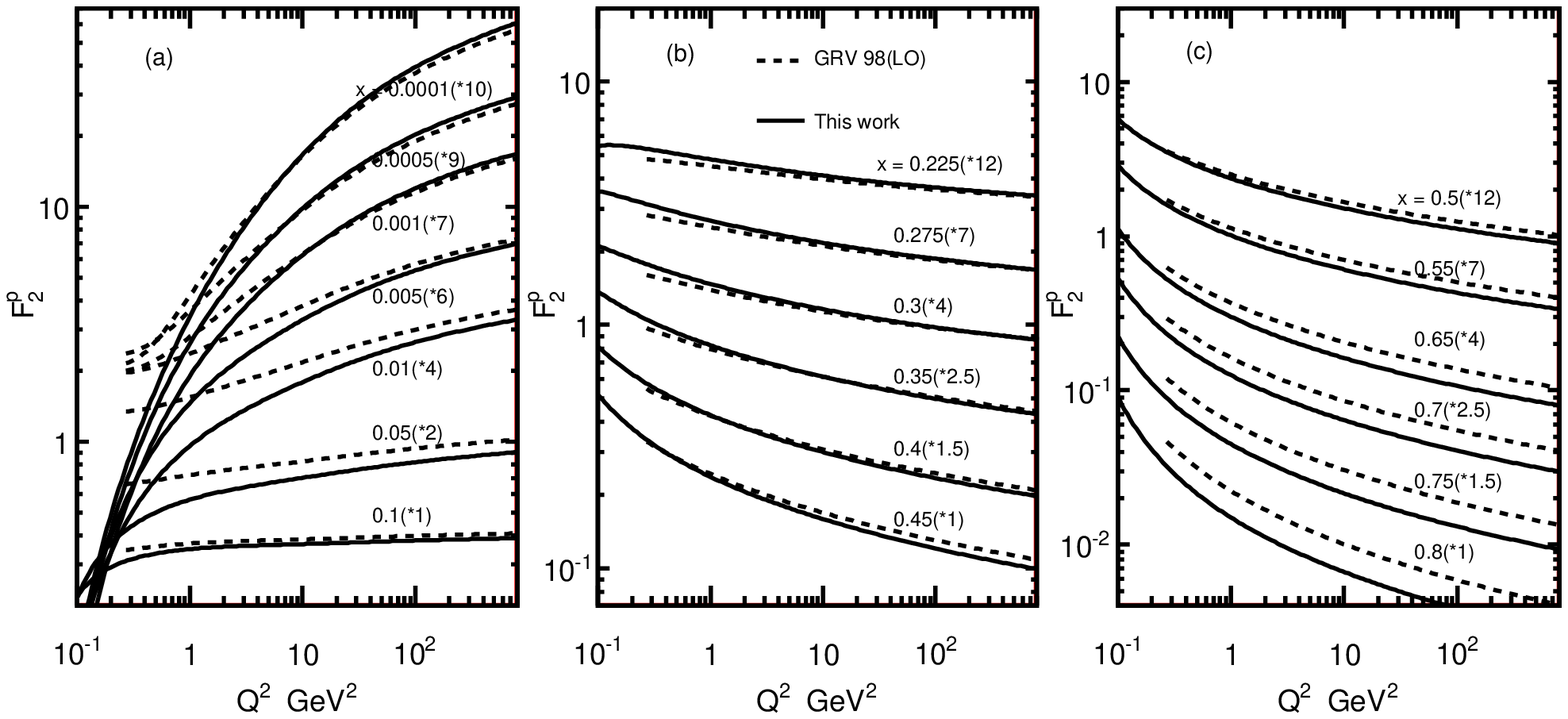}
\caption{Comparisons of the $Q^2$-dependence of our predicted $F_2^P$ (solid curves)
with the GRV(98LO) results (dashed curves) \cite{25}.
For a better display, the structure
function values are scaled  at each $x$ by the factors shown in
brackets.}
\label{fig8}
\end{figure}

\begin{figure}[htp]
\centering
\includegraphics[width=0.5\textwidth]{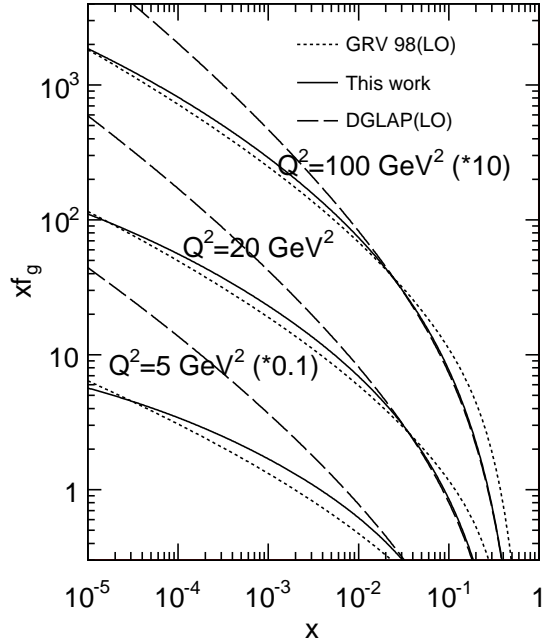}
\caption{Comparisons among our predicted gluon distribution (solid curves), the
GRV(98LO) (dashed curves) \cite{25} and the results using the DGLAP evolution with the natural input (broken curves).}
\label{fig9}
\end{figure}

\begin{figure}[htp]
\centering
\includegraphics[width=0.5\textwidth]{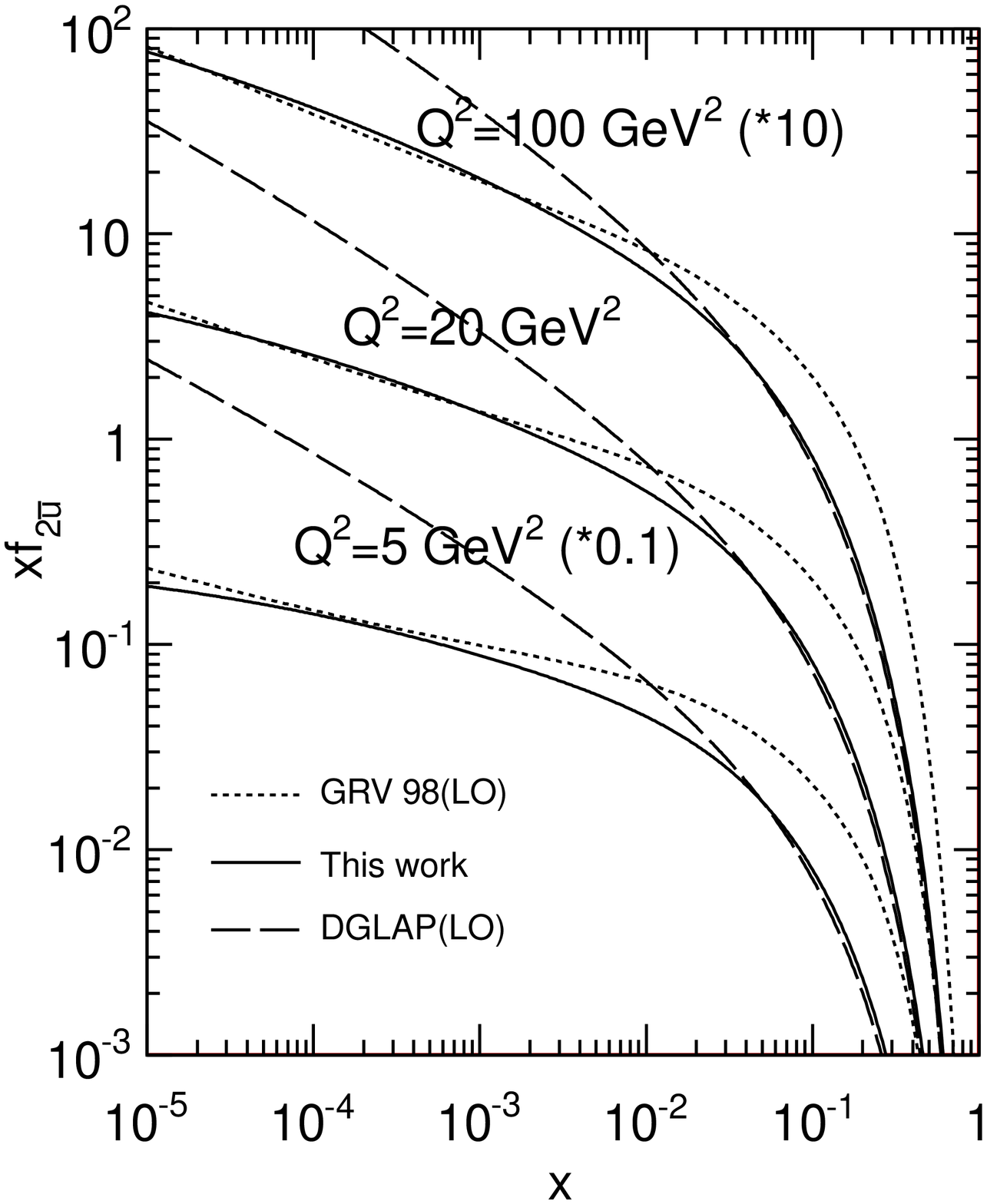}
\caption{Similar to Fig. 9, but for the sea quark distributions.}
\label{fig10}
\end{figure}

    We try to explore the parton distributions down to $Q^2< 1~$GeV$^2$ at small $x$
using Eqs. (1-3). Our predicted proton structure functions
are shown in Fig. 11, where we plot the contributions of the Regge
part in the Donnachie-Landshoff (DL) model \cite{60}. The prediction
of gluon and sea quark distributions at low $Q^2$ based on Eqs. (1-3) is presented in Fig. 12. Although the predicted
structure functions are lower than the ZEUS data \cite{52,53,54,55,56}, the
plateau-like shapes allow us to establish a smooth connection
between the partonic and non-partonic pictures of the nucleon's
structure functions in low $Q^2$ and small $x$ range \cite{61}. It
will be studied in our following work.

\begin{figure}[htp]
\centering
\includegraphics[width=0.7\textwidth]{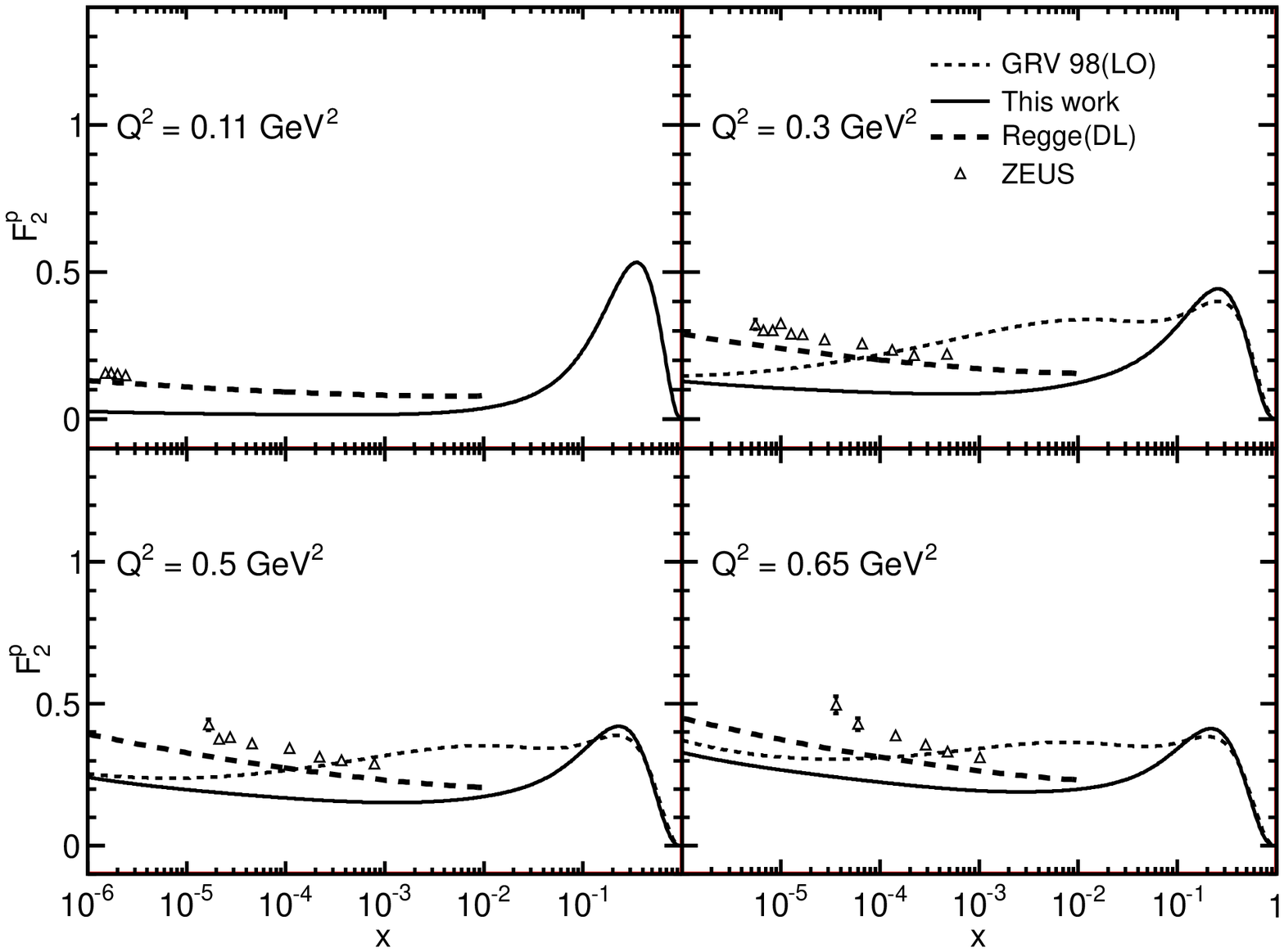}
\caption{Our predicted proton structure function (solid curves) in low $Q^2$
range. The contributions of Regge part in the Donnachie-Landshoff
model (broken curves) \cite{60}, GRV(98LO)\cite{25} results (dashed curves) and the ZEUS data \cite{52,53,54,55,56} are presented.}
\label{fig11}
\end{figure}

\begin{figure}[htp]
\centering
\includegraphics[width=0.6\textwidth]{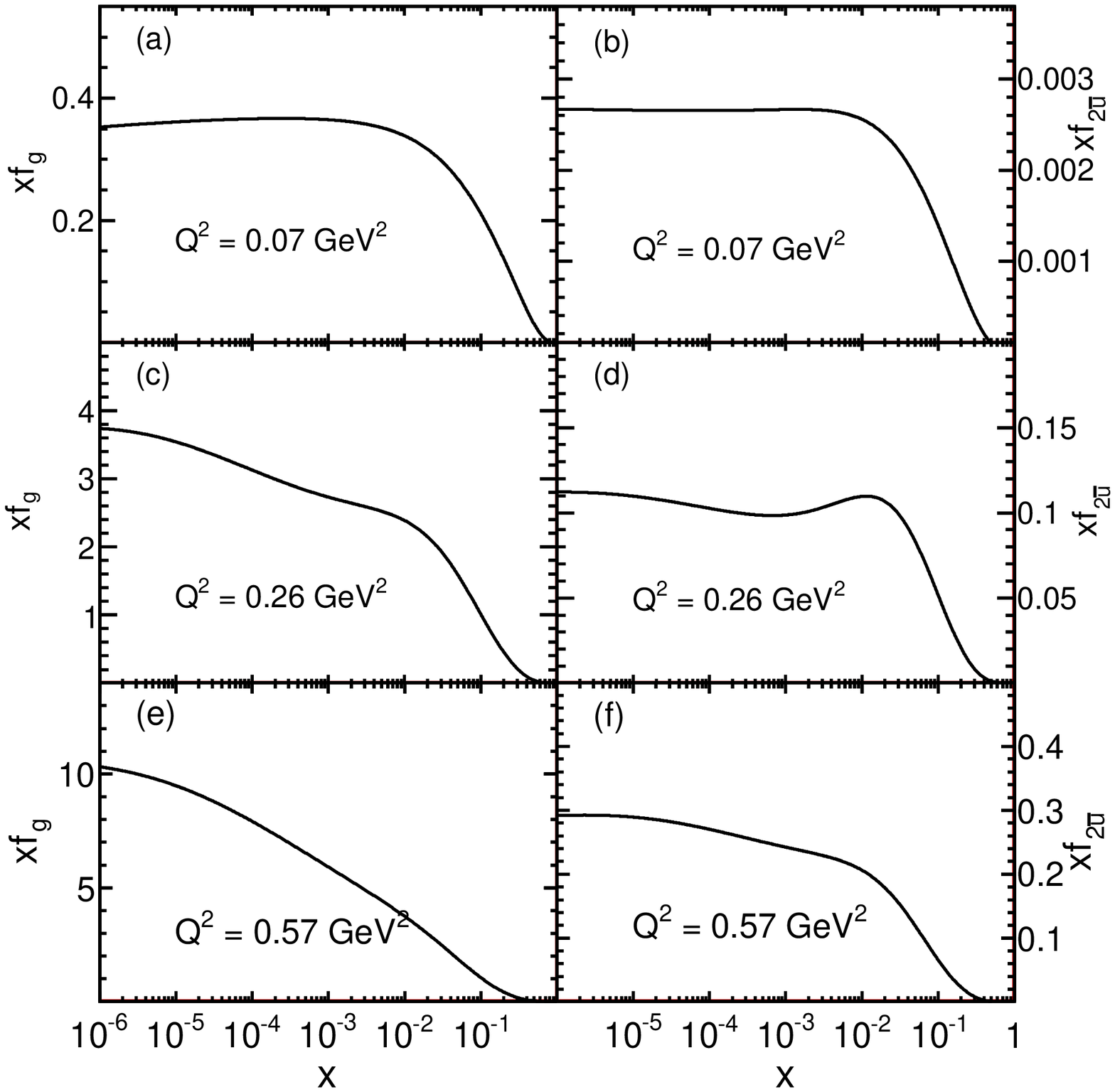}
\caption{Our predicted gluon (Left) and sea quarks (Right) distributions in the low $Q^2$ range.}
\label{fig12}
\end{figure}

      The difference between our model and GRV model becomes obvious near the GRV starting
point $Q^2\simeq 0.26~ $GeV$^2$ (Fig. 13). The comparisons of our model with the
GRV and CJ12 \cite{62} databases at $Q^2=1.4~$GeV$^2$ are presented in
Fig. 14.

\begin{figure}[htp]
\centering
\includegraphics[width=0.6\textwidth]{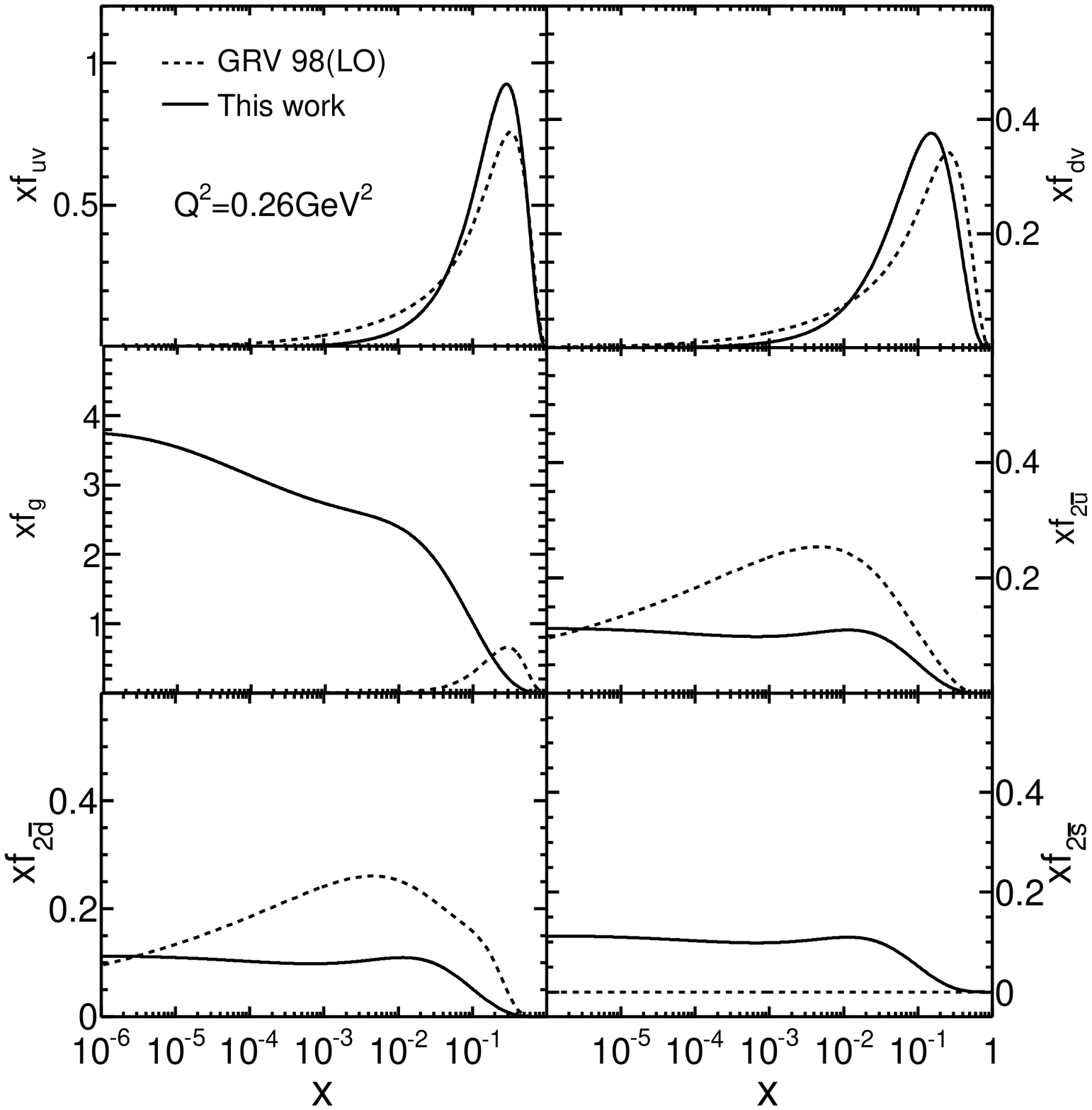}
\caption{Comparisons of our predicted parton distributions (solid curves) with the GRV input (dashed curves)
at the GRV-starting point $Q^2=0.26 $ GeV$^2$.}
\label{fig13}
\end{figure}

\begin{figure}[htp]
\centering
\includegraphics[width=0.6\textwidth]{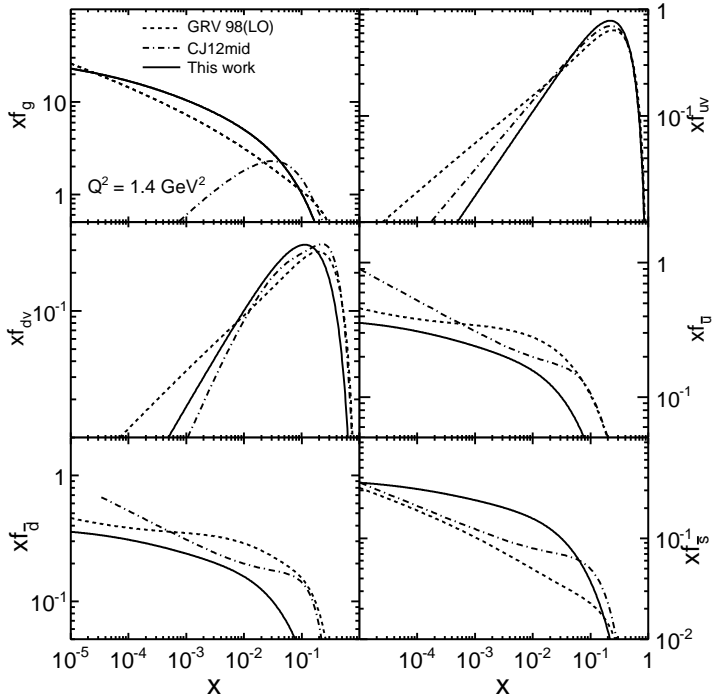}
\caption{Comparisons of our predicted parton distributions (solid curves) with
GRV98\cite{25}and CJ12\cite{62}
distributions at $Q^2=1.4 $ GeV$^2$.}
\label{fig14}
\end{figure}

    The predicted gluon distributions of the different databases exhibit large difference.
The reason is that the lepton probes can not directly measure the
gluon distribution. The parameters of input gluon distribution are
determined by higher order QCD processes, such as the
scaling violation and longitudinal structure function $F_L$. Because only limited  amount of data are available,
 the constraints on the input gluon distribution are much
looser than for quarks. Different from those global databases, both
gluon and sea quarks are dynamically generated from the input
valence quarks based on Eqs. (1-3) in this work. Once
the valence quark input is fixed by the observed quark
distributions, the gluon distribution is also determined in the
leading order. The comparisons of the $x$-dependence of the gluon
distributions at given $Q^2$ are plotted in Figs. 15-17.

\begin{figure}[htp]
\centering
\includegraphics[width=0.7\textwidth]{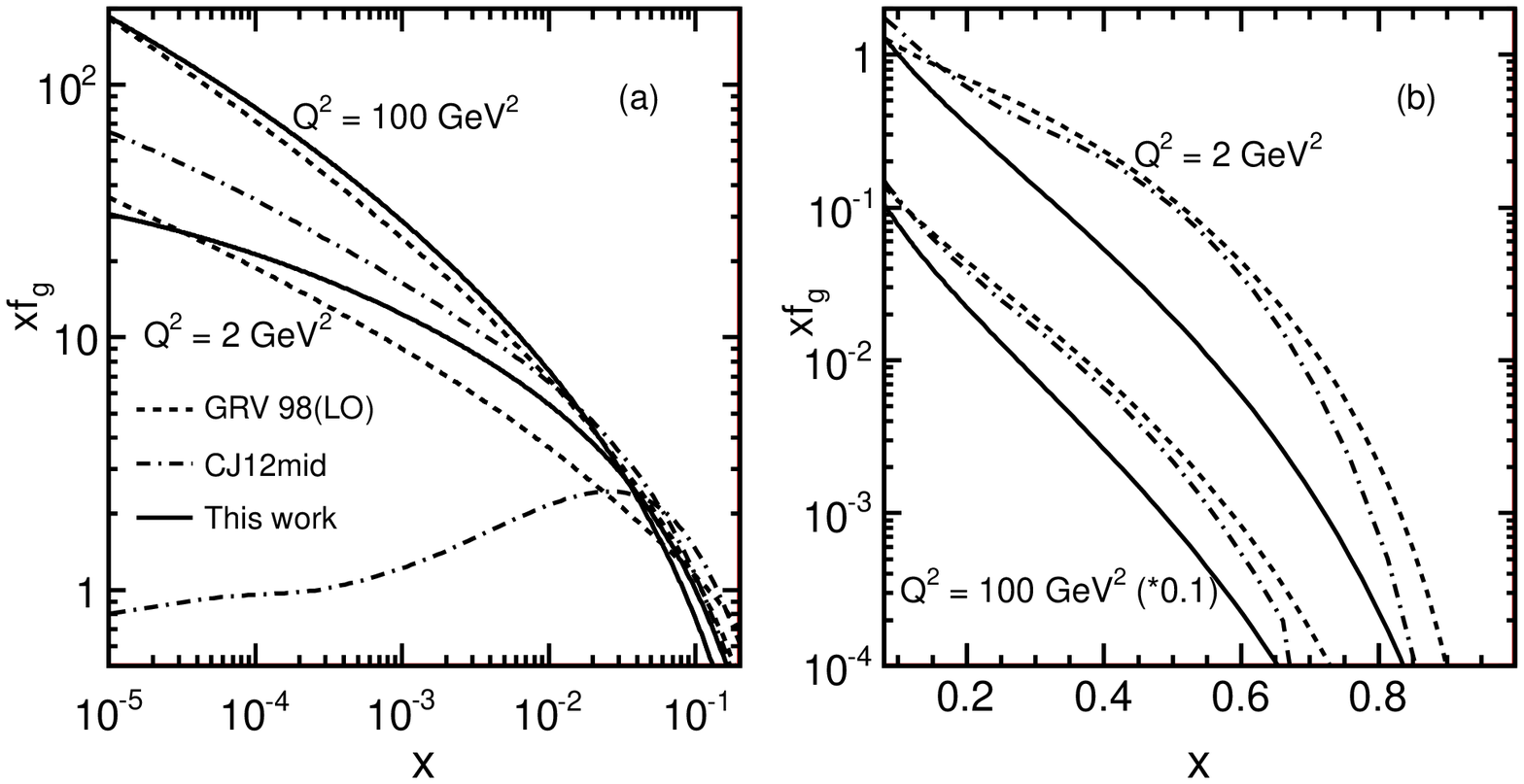}
\caption{Comparisons of our predicted $x$-dependence of gluon distribution (solid curves)
with GRV98\cite{25} and CJ12\cite{62} distributions
at different $Q^2$.}
\label{fig15}
\end{figure}

\begin{figure}[htp]
\centering
\includegraphics[width=0.6\textwidth]{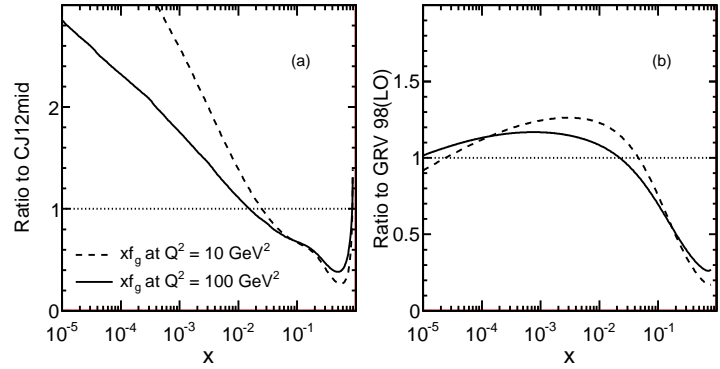}
\caption{Ratios of our predicted gluon distribution to (a) CJ12\cite{62} and (b) GRV98\cite{25} distributions at $Q^2$ = 10 and 100 GeV$^2$.}
\label{fig16}
\end{figure}

\begin{figure}[htp]
\centering
\includegraphics[width=0.5\textwidth]{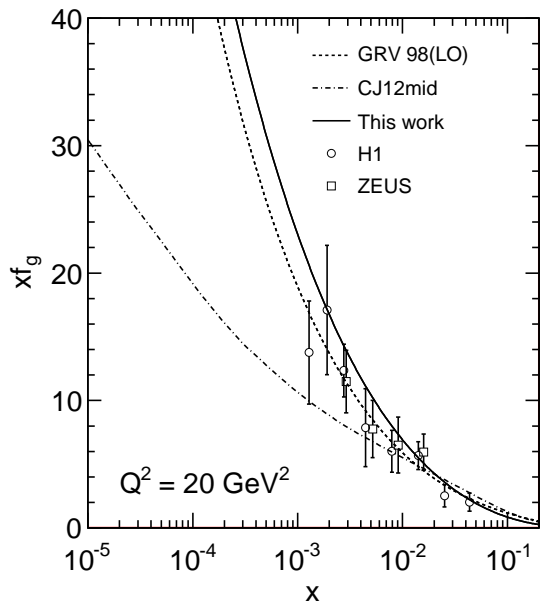}
\caption{Comparisons of our predicted gluon distribution at
$Q^2=20 $ GeV$^2$ with the HERA data. The results of GRV98\cite{25} and CJ12\cite{62} gluon distributions also are presented.}
\label{fig17}
\end{figure}

\newpage
\begin{center}
\section{Discussions and Summary}
\end{center}

    An unavoidable question is whether Eq. (1) can be used at
$Q^2 ~ \ge ~0.064~$GeV$^2$. Now let us discuss this question.

     One may think that a large value of $\alpha_s$ at low $Q^2$ will lead to a divergent perturbative expansion.
Refs. \cite{7,8,9} have emphasized that the DGLAP evolution is still in the
perturbative region even at low static point $\mu^2\sim 0.064~
$GeV$^2$ since the expansion factor is $\alpha_s(\mu^2)/2\pi <1 $ in the
DGLAP equation and the evolution kernels are non-singular at low
$Q^2$. The GRV model was extended to the next leading order and even to the
next-next leading order approximation at $Q^2<1~$GeV$^2$ \cite{63}. These results
indicate that the higher order corrections to the DGLAP evolution
are small.

    We show that these conclusions are still
valid for the DGLAP equation with the ZRS corrections. The
perturbative approximation may be invalid if the net (shadowing and
antishadowing) effect is positive. As we have pointed out that the
size of the ZRS correction is not only dependent on the
magnitude of $\alpha_s(Q^2)$, but also is related to the shape of
the parton densities in the range of $x$ to $x/2$ \cite{65}. The net
nonlinear correction will be positive if the parton distribution
$\sim x^{-\lambda}$ ($\lambda>\lambda_{BFKL}\sim 0.5$). However, the
radioactively generated partons by the DGLAP evolution at small $x$
behave as $\sim x^{-\lambda}$ ($\lambda<\lambda_{BFKL}$) even
through it is steep. In this case the shadowing effect in the
evolution process will be weakened by the antishadowing effect, and
the net effect of the parton fusions always keeps the negative
correction. The parton distributions will asymptotically approach
a finite value by the action of the net shadowing in the leading
order level. On the other hand, the contributions of the higher
order recombination (i.e., the multi-parton recombination) is
negligible since the parton densities are low at $Q^2< 1~$GeV$^2$.
We also notice the possibility that the nonperturbative dynamics of
QCD generates an effective gluon mass at very low $Q^2$ region
\cite{35}, or gluons become Abelian gluons when $Q^2\rightarrow
\Lambda^2_{QCD}$ \cite{69,70}. A simple phenomenological approach can be
used to estimate the above corrections to the evolution of the
parton distributions: the QCD running coupling constant is frozen at
an infrared value, i.e., $\alpha_s(Q^2)\le \kappa$, $\kappa$ is an
undetermined parameter.  Using this restriction we obtain
the similar results after adjusting the input distributions. Our
numerical calculations also show that the resummations
$\sum_n[\alpha_s/(2\pi)\ln(Q^2/\Lambda_{QCD})]^n$ and
$\sum_n[\alpha_s^2/(4\pi R^2Q^2)]^n$ converge quickly, i.e., the
perturbative evolution of Eq. (1) is stable at low $Q^2$.

        Our analysis has shown that the parton recombination plays an important role at small $x$ and low $Q^2$.
Although any power law correction including parton recombination will
disappear at high $Q^2$, the contributions of the nonlinear terms in
Eq.(1) at low $Q^2$ will be "remembered" in the parton distribution
evolution process and observed at high $Q^2$. For example, with the
same natural input distribution there is a big difference between
the parton distributions from the DGLAP equation and the ones from
Eqs. (1-3), as shown in Figs. 9 and 10.

    The input valence quark distributions parameterize
the interactions of constituents of the proton at scale
$Q^2\rightarrow \Lambda_{QCD}$ in the strong coupling regime. One
can use chiral effective field theories to model the input
valence quark distributions as Ref. \cite{27}. It is interesting to
notice that these distributions are compatible with our
parameterized input shown in Fig. 18. Therefore, we think that the
DGLAP equation with the ZRS corrections is a powerful tool to connect the effective QCD
quark model of hadron at scale $\mu^2$ with the observed structure
function at high scale $Q^2>> \mu^2$.

\begin{figure}[htp]
\centering
\includegraphics[width=0.6\textwidth]{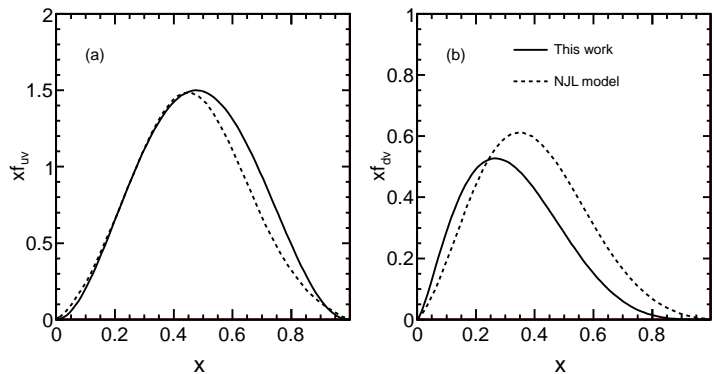}
\caption{Comparisons of our input quark distributions (solid curves) with the
valence quark distributions in the Nambu-Jona-Lasinio model (dashed curves) \cite{27}.}
\label{fig18}
\end{figure}

           The parton distributions of the proton in this work can be generalized to
nuclear target, where the nuclear shadowing effect is a natural
result of the parton recombination among different bound nucleons,
while the nuclear environment deforms the input valence quarks. We will
discuss them in our next work.

    In this work, all sea quarks are generated perturbatively by gluon
splitting. Therefore, the sea quark distributions have isospin
symmetry. However, experiments questioned these naive expectations
\cite{71,72}. These effects require nonperturbative explanations. In
this respect, the DGLAP equation with ZRS corrections provides an
effective way to test various models which may modify the input
distributions. Furthermore, in our following works, the results of
this work will be used to dynamically predict the nuclear gluon
distributions and to reduce the free parameters related to EMC
effects.

     In summary, the parton distributions in the proton are
evaluated dynamically starting from three valence quarks input at
the low scale $\mu^2$ by the DGLAP equation with the ZRS corrections. Our results show
that negative nonlinear corrections improve the perturbative
stability of the QCD evolution equation at low $Q^2$. Our predicted
parton distributions of the proton with four free parameters are
compatible with the existing databases.
 We show that the sea quark distributions
exhibit a positive and plateau-like behavior at small $x$ and low $Q^2$.
This approach provides a powerful tool to connect the
quark models of the hadron and various non-perturbative effects at the scale $\mu^2$ with the measured
structure functions at the high scale $Q^2>> \mu^2$.

\vspace{0.3cm}

\noindent {\bf Acknowledgments}: This work is partly supported by the National Natural Science Foundations of China under the Grants
Number 10875044, 11275120 and Century Program of
Chinese Academy of Sciences Y101020BR0.

\newpage
\noindent {\bf Appendix A. The GLR-MQ corrections.}

    For comparing with the ZSR corrections, we present the GLR-MQ corrections to the DGLAP
equation at the DLL approximation [9]

$$Q^2\frac{dxf_g(x,Q^2)}{dQ^2}$$
$$=Q^2\frac{dxf_g(x,Q^2)}{dQ^2}\vert_{DGLAP}-\frac{9\alpha_s^2(Q^2)}{2R^2Q^2}\int_x^{10^{-2}}\frac{dy}{y}
[ yf_g(y,Q^2)]^2,\eqno(A-1)$$

$$Q^2\frac{dxf_{\overline{q}_i}(x,Q^2)}{dQ^2}$$
$$=Q^2\frac{dxf_{\overline{q}_i}(x,Q^2)}{dQ^2}\vert_{DGLAP}-\frac{3\alpha_s^2(Q^2)}{20R^2Q^2}
[ xg(x,Q^2)]^2$$
$$ +\frac{6\alpha_s}{\pi Q^2}\int_x^{10^{-2}}\frac{dy}{y}\frac{-2y^3x^2+15y^2x^3-30yx^4+18x^5}{y^5}
yG_{HT}(y,Q^2)\eqno(A-2), $$ where

$$Q^2\frac{dyG_{HT}(y,Q^2)}{dQ^2}=-\frac{9\alpha_s^2}{2R^2}\int_y^{10^{-2}}\frac{dz}{z}[zG(z,Q^2)]^2\eqno(A-3).$$


\newpage
\begin{thebibliography}{99}

\bibitem{1} D.J. Gross and F. Wilczek, \textit {Phys. Rev. D} $\bf {8}$ (1973) 3633.


\bibitem{2} D.J. Gross and F. Wilczek, \textit {Phys. Rev. D} $\bf {9}$ (1974) 980.


\bibitem{3} H. Georgi and H.D. Politzer, \textit {Phys. Rev. D} $\bf {9}$ (1974) 416.

\bibitem{4} G. Altarelli and G. Parisi, \textit{Nucl. Phys. B} $\bf{126}$ (1977) 298.

\bibitem{5} V.N.Gribovand L.N. Lipatov, \textit{Sov. J. Nucl. Phys.} $\bf{15}$(1972) 438.

\bibitem{6} Yu.L.Dokshitzer, \textit{Sov. Phys. JETP.} $\bf{46}$ (1977) 641.

\bibitem{7} G. Parisi and R. Petronzio, \textit{Phys. Lett. B} $\bf{62}$ (1976) 331.

\bibitem{8} V.A. Novikov, M.A. Shifman, A.I. Vainshtein, V.I. Zakharov, \textit{JETP Lett.} $\bf{24}$ (1976) 341.

\bibitem{9} M. Gl\"{u}ck, E. Reya, \textit{Nucl. Phys. B} $\bf{130}$ (1977) 76.

\bibitem{10} R. L. Jaffe, \textit{Phys. Rev. D} $\bf{11}$ (1975) 1953.

\bibitem{11} A. Le Yaouanc, L. Oliver, O. Pene, and J. C. Raynal, \textit{ Phys. Rev. D} $\bf{11}$
(1975) 1272.

\bibitem{12} R.J. Hughes, \textit{Phys. Rev. D} $\bf{16}$ (1977) 622.


\bibitem{13} J.S. Bell and A.J.G. Hey, \textit{Phys. Lett.B} $\bf{74}$ (1978) 77.


\bibitem{14} C.J. Benesh and G.A. Miller, \textit{Phys. Rev. D} $\bf{36}$ (1987) 1344.


\bibitem{15} I.C. Cloet, W. Bentz, A.W. Thomas, \textit{Phys. Lett. B}$\bf{621}$ (2005) 246.

\bibitem{16}  M. Gl\"{u}ck, E. Reya, A. Vogt, \textit{Z. Phys. C} $\bf{48}$ (1990) 471.

\bibitem{17} L.V. Gribov, E.M. Levin and M.G. Ryskin, \textit{Phys. Rep.} $\bf{100}$ (1983) 1.

\bibitem{18} A.H. Mueller and J. Qiu, \textit{Nucl. Phys. B} $\bf{268}$ (1986) 427.

\bibitem{19} J.C. Collins and J. Kwiecinski, \textit{Nucl. Phys. B} $\bf {335}$ (1990) 89.

\bibitem{20} J. Bartels, G.A. Schuler and J. Blumlein, \textit{Z. Phys. C} $\bf {50}$ (1991) 91.

\bibitem{21} M. Altmann, M. Gl\"uck and E. Reya, \textit{Phys. Lett. B}
 $\bf {285}$ (1992) 359.

\bibitem{22} W. Zhu, \textit{Nucl. Phys. B} $\bf{551}$ (1999) 245 [arXiv:hep-ph/9809391];

\bibitem{23} W. Zhu, J.H. Ruan, \textit{Nucl. Phys.B} {\bf 559} (1999) 378
[arXiv:hep-ph/9907330v2];

\bibitem{24} W. Zhu and Z.Q. Shen, \textit{ HEP $\&$ NP,} $\bf{29}$ (2005) 109 [arXiv:hep-ph/0406213v3].

\bibitem{25}  M. Gl\"{u}ck, E. Reya, and A. Vogt, \textit{Eur. Phys.
J.C} $\bf{5}$ (1998) 461.

\bibitem{26} P.D.B. Collins, An Introduction to Regge Theory and High Energy Scattering, \textit{Cambridge University Press } (1977).

\bibitem{27} H. Mineo, W. Bentz, N. Ishii, A.W. Thomas, K. Yazaki, \textit{Nucl. Phys. A} $\bf{735}$ (2004) 482.

\bibitem{28} V.A. Abramovsky, J.N. Gribov and O.V. Kancheli, \textit{Sov. J. Nucl. Phys.} $\bf{18}$ (1974) 308.

\bibitem{29} W. Zhu, D.L. Xue, K.M. Chai and Z.X. Xu, \textit{Phys. Lett. B} $\bf{317}$ (1993) 200.

\bibitem{30} J.C. Collins and J.W. Qiu, \textit{Phys. Rev.D} $\bf{39}$ (1989) 1398.

\bibitem{31} W. Zhu, \textit{Chin. Phys. Lett.} $\bf{16}$ (1999) 481.

\bibitem{32} C.F. von Weizs\"{a}ker, \textit{Z. Phys.} $\bf{88}$ (1934) 612.

\bibitem{33} E.J. Williams, \textit{Phys. Rev. }$\bf{45}$ (1934) 729.

\bibitem{34} V.N Baier, V.S. Fadin and V.A. Khoze, \textit{Nucl. Phys. B} $\bf{65}$ (1973) 381.

\bibitem{35} M.S. Chen and P. Zerwas, \textit{Phys. Rev. D} $\bf{12}$ (1975) 187.

\bibitem{36} L.N. Lipatov, \textit{Sov. J. Nucl. Phys.} $\bf{23}$ (1976) 338.

\bibitem{37} V.S. Fadin, E.A. Kuraev, L.N. Lipatov, \textit{Phys. Lett. B} $\bf{60}$ (1975) 50.

\bibitem{38} E.A. Kuraev, L.N. Lipatov, V.S. Fadin, \textit{ Sov. Phys. JETP} $\bf{44}$ (1976) 443.

\bibitem{39} E.A. Kuraev, L.N. Lipatov, V.S. Fadin, \textit{Sov. Phys. JETP} $\bf{45}$ (1977) 199.

\bibitem{40} I.I. Balitsky, L.N. Lipatov, \textit{Sov. J. Nucl. Phys. } $\bf{28}$ (1978) 822.

\bibitem{41} I.I. Balitsky, L.N. Lipatov, \textit{JETP Lett. } $\bf{30}$ (1979) 355.

\bibitem{42} W. Zhu, \textit{Phys. Lett.} $\bf{B389}$ (1996) 374.

\bibitem{43} W.Zhu, J.H. Ruan, J.F. Yang and Z.Q. Shen, \textit{Phys.Rev.D} $\bf{68}$ (2003) 094015.

\bibitem{44} J.H. Ruan and W. Zhu, \textit{ Phys.Rev.C} $\bf{80}$ (2009) 045209.

\bibitem{45} J.H. Ruan and W. Zhu, \textit{Phys. Rev.C} $\bf{81}$ (2010) 055210.

\bibitem{46} W. Zhu, J.H. Ruan and F.Y. Hou, \textit{Int. J. Mod. Phys. E } $\bf{22}$ (2013) 1350013.

\bibitem{47} H1 Collab. (S. Aid \textit{et al.}), \textit{ Nucl. Phys.B} $\bf{470}$ (1996) 3.

\bibitem{48} H1 Collab. (C. Adloff \textit{et al.}), \textit{Nucl. Phys. B} $\bf{497}$ (1997) 3.

\bibitem{49} H1 Collab. (C. Adloff \textit{et al.}), \textit{Eur. Phys. J.C} $\bf{21}$ (2001) 33 [arXiv:hep-ex/0012053].

\bibitem{50} H1 Collab. (C.Adloff \textit{et al.}), \textit{Eur. Phys. J. C} $\bf{13}$ (2000) 609
[arXiv:hep-ex/9908059].

\bibitem{51} H1 Collab. (C. Adloff\textit{et al.}), \textit{Eur. Phys. J. C} $\bf{19}$ (2001) 269 [arXiv:hep-ex/0012052].

\bibitem{52} ZEUS Collab. (M. Derrick \textit{et al.}), \textit{Z. Phys. C} $\bf{69}$ (1996) 607.

\bibitem{53} ZEUS Collab. (M. Derrick \textit{et al.}), \textit{Z. Phys. C} $\bf{72}$ (1996) 399.

\bibitem{54} ZEUS Collab. (S. Chekanov \textit{et al.}), \textit{Eur. Phys. J. C} $\bf{21}$ (2001) 443 [arXiv:hep-ex/0105090].

\bibitem{55} ZEUS Collab. (J.Breitweg \textit{et al.}), \textit{Phys. Lett. B} $\bf{487}$ (2000) 53 [arXiv:hep-ex/0005018].

\bibitem{56} ZEUS Collab. (J. Breitweg \textit{et al.}), \textit{Phys. Lett. B}$\bf{407}$ (1997) 432.

\bibitem{57} The New Muon Collab. (M. Arneodo \textit{et al.}), \textit{Nucl. Phys. B} $\bf{483}$ (1997)
3 [arXiv:hep-ph/9610231].

\bibitem{58} BCDMS Collab. (A. C. Benvenuti \textit{et al.}), \textit{Phys. Lett. B} $\bf{223}$ (1989) 485 .

\bibitem{59} E665 Collab. (M. R. Adams \textit{et al.}), \textit{Phys. Rev. D } $\bf{54}$ (1996) 3006 .

\bibitem{60} A. Donnachie and P.V. Landshoff, \textit{Phys. Lett. B} $\bf{437}$ (1998) 408.

\bibitem{61} for see A. De Roeck, R.S. Thorne,  Prog. Part. \textit{Nucl. Phys.} $\bf{66}$ (2011) 727 [arXiv:1103.0555].

\bibitem{62} J.F. Owens, \textit{Phys. Rev. D } $\bf{87}$ (2013) 094012 [arXiv:1212.1702].

\bibitem{63} M. Gl\"{u}ck, C. Pisano, E. Reya, \textit{Eur. Phys. J.C} $\bf{40}$ (2005) 515;

\bibitem{64} P. Jimenez-Delgado, E. Reya, \textit{Phys. Rev. D} $\bf{79}$ (2009) 074023.

\bibitem{65} W. Zhu, K.M. Chai and B. He, \textit{ Nucl. Phys. B} $\bf{427}$ (1994) 525.

\bibitem{66} W. Zhu, K.M. Chai and B. He,\textit{Nucl. Phys.B} $\bf{449}$ (1995) 183.

\bibitem{67} E.G.S. Luna, \textit{Braz. J. Phys.} $\bf {37}$ (2007) 84.

\bibitem{68} A.A. Natale, \textit{Braz. J. Phys.} $\bf{37}$ (2007) 306.

\bibitem{69} Y. M. Cho, \textit{Phys. Rev. Lett.} $\bf{46}$ (1981) 302;

\bibitem{70} Y. M. Cho, \textit{Phys. Rev. D} $\bf{23}$ (1981) 2415.

\bibitem{71} NA51 Collab. (A. Baldit \textit{et al.}), \textit{Phys. Lett. B} $\bf{B332}$, 244 (1994);

\bibitem{72} FermilabE866/NuSea Collab. (E.A. Hawker \textit{et al.}), \textit{Phys. Rev. Lett.} $\bf{80}$ (1998) 3715.


\end{thebibliography}
\end{document}